\documentclass[aps,pre,twocolumn,amssymb,amsmath,amsfonts,floatfix,groupedaddress,citeautoscript]{revtex4-2}
\usepackage{hyperref}
\usepackage{graphicx}
\begin{document}

\title{Percolation and electrical conduction in random systems of curved linear objects on a plane: computer simulations along with a mean-field approach}%

\author{Yuri~Yu.~Tarasevich}
\email[Corresponding author: ]{tarasevich@asu.edu.ru}

\author{Andrei~V.~Eserkepov}
\email{dantealigjery49@gmail.com}

\author{Irina~V.~Vodolazskaya}
\email{vodolazskaya\_agu@mail.ru}

\affiliation{Laboratory of Mathematical Modeling, Astrakhan State University, Astrakhan, 414056, Russia}

\date{\today}
\begin{abstract}
Using computer simulations, we have studied the percolation and the electrical conductance of two-dimensional, random percolating networks of curved, zero-width metallic nanowires. We mimicked the curved nanowires using circular arcs. The percolation threshold decreased as the aspect ratio of the arcs increased. Comparison with published data on the percolation threshold of  symmetric quadratic B\'{e}zier curves suggests that, when the percolation of slightly curved wires is simulated, the particular choice of curve to mimic the shape of real-world wires is of little importance. Considering the electrical properties, we took into account both the nanowire resistance per unit length and the junction (nanowire/nanowire contact) resistance. Using a mean-field approximation (MFA), we derived the total electrical conductance  of the nanowire-based networks as a function of their geometrical and physical parameters. The MFA predictions have been confirmed by our Monte Carlo numerical simulations. For our  random homogeneous and isotropic systems of conductive curved wires, the electric conductance decreased as the wire shape changed from a stick to a ring when the wire length remained fixed.
\end{abstract}

\pacs{}
\maketitle

\section{Introduction}\label{sec:intro}
Nanowire-based transparent conductive films (TCFs) or transparent conductive electrodes (TCEs) are promising candidates being considered to replace traditional indium tin oxide (ITO) devices~\cite{Gao2016,McCoul2016,Zhang2020}. Nanowires (NWs) are often curved~\cite{Yin2015,Yin2017}. For instance, a curvature distribution for silver nanowires (AgNWs) has been reported~\cite{Kang2018}.

The effects of perturbations in length, angular anisotropy, and the radius of curvature of one-dimensional (1D) objects on the onset of percolation have previously been studied~\cite{Langley2018}. The authors reported that wire curvatures have a small impact on the network, resulting in a moderate increase in the percolation threshold.

The percolation threshold of curved 1D objects, described as symmetric quadratic B\'{e}zier curves, has also  been investigated~\cite{Lee2021}. Here, eight different values of the curvature were used to study the dependency of the percolation threshold on the curvature. For each value of curvature, the electrical conductivity of the system under consideration was calculated using the junction resistance dominant assumption, viz., the ratio of the junction resistance to the resistance of the wire was taken as $10^3$. The number density of the wires
\begin{equation}\label{eq:n}
n = \frac{N}{L^2}.
\end{equation}
was changed in the range $0.1 \leqslant n/n_\text{c} \leqslant 1$, where $n_\text{c}$ is the critical number density (the percolation threshold). The apparent conductivity exponents
(see, e.g., Refs.~\cite{Li2010,Zezelj2012,Lee2021})
\begin{equation}\label{eq:t}
  t = \frac{d \log \sigma}{ d \log (n - n_\text{c})}
\end{equation}
were estimated to be in the range of $\tilde{t}= 1.40-1.42$.

The percolation of randomly placed and oriented arcs of unit length in an $L \times L$ square matrix has previously been studied using  Monte Carlo (MC) simulation~\cite{Lin2010}. The authors found that the percolation threshold increased with increasing curl ratio. The curl ratio was defined as the ratio between the length and the largest distance of two arbitrary points on the wire. That definition seems to contain a misprint, since the widely used definition says that the curl ratio is the ratio of the curve length divided by the end-to-end distance (see, e.g., Refs.~\onlinecite{Yi2004,Fata2020}). In this case, for a circular arc, the curl ratio is
\begin{equation}\label{eq:CR}
  \varkappa = \frac{\alpha}{2\sin\frac{\alpha}{2}},
\end{equation}
where $\alpha$ is the central angle of the arc.  The percolation threshold was determined as the critical volume fraction
\begin{equation}\label{eq:CVF}
  \phi_\text{c} = \frac{N_\text{c} l^2_\text{w}}{\epsilon_\text{w} L^2},
\end{equation}
where $\epsilon_\text{w}$ is the aspect ratio of the wires (was set as 1000 for all simulations in Ref.~\onlinecite{Lin2010}) and $l_\text{w}$ is the wire length ($l_\text{w}=1$ in Ref.~\onlinecite{Lin2010}). Since,  in Ref.~\onlinecite{Lin2010}, the critical volume fraction is presented as a percentage, the value of the critical number density is simply $n_\text{c} = 10\phi_\text{c}$.

MC simulations have been employed to compute `the curviness percolation threshold' in two-dimensional (2D) networks consisting of curvy nanotubes or nanowires~\cite{Wang2021}. Third order B\'{e}zier curves were used to simulate these curvy nanotubes/nanowires. This study showed that nanotube/nanowire curviness plays a significant role in determining the percolation threshold. Hybrid network structures composed of straight and curved nanowires modeled as semi-circular wires have been studied, and the effects caused by the curvature on the overall conductance response of the film were examined~\cite{Esteki2021}.

The effect of nanowire curviness on the percolation resistivity of metal nanowire-based networks has also been studied~\cite{Hicks2018,Fata2020}. Curved NWs were modeled using third-order B\'{e}zier curves. The network conductivity exhibited an inverse power law dependence on the curl ratio.

Recently, mean-field approximation (MFA) has been successfully applied to random dense 2D systems composed of nanorings~\cite{Tarasevich2021,Tarasevich2022b}, nanosticks~\cite{Tarasevich2022,Tarasevich2022a} and their mixtures~\cite{Tarasevich2022c}. The goal of the present study is the application of MFA to dense 2D systems consisting of randomly deposited conductive curved nanowires. Circular arcs seem to be a natural choice to mimic the shape of such curved NWs. The percolation threshold in random 2D systems of curved 1D objects is expected to be dependent both on the curvature and on the particular shape of the objects. We intend to shed light on this issue as well.

The rest of the paper is constructed as follows. Section~\ref{sec:methods} presents our computational and analytical methods, viz., Sec.~\ref{subsec:simul} describes some technical details of our simulation, Sec.~\ref{subsec:analytics} is devoted to an analytical consideration using a MFA. In Section~\ref{sec:results}, we present our main results and compare the MFA predictions with computer simulations. Section~\ref{sec:concl} summarizes the main results and suggests possible directions for further study.

\section{Methods\label{sec:methods}}
\subsection{Sampling and main definitions\label{subsec:sampling}}
Consider a domain of size $L \times L$ in which $N$ metallic curved NWs are randomly placed and equiprobably oriented. We mimic the shape of any curved NW by an arc of a circle of radius $r$. The arc rests on the central angle $\alpha$. When the arc length, $l_\text{w}$, is fixed, while the curvature radius tends to infinity, $r \to \infty$, and thus the arc tends to a linear segment ($\alpha = 0$). Another limiting case, when $r = l/(2\pi)$, corresponds to a circle ($\alpha = 2\pi$). The most precise known values of the percolation thresholds are $n_\text{c}l^2 = 5.6372858(6)$ and $n_\text{c}l^2 = 1.43632545(8)$ for sticks and discs, respectively~\cite{Mertens2012}. Here, $l$ means the stick length or the disc diameter, i.e., the maximal dimension of the particle. The percolation threshold of arcs is expected to change between these two boundaries, when the central angle, $\alpha$, varies from 0 to~$2\pi$.

Since the curvature of the B\'{e}zier curve changes along the curve, using this quantity is hardly suitable for comparing the percolation thresholds of systems composed from circular arcs with systems composed from B\'{e}zier curves. By contrast, the aspect ratio does seem to be an appropriate quantity for such a comparison. In the case of arcs, the maximal dimension is the chord length ($C = 2 r \sin \frac{\alpha}{2}$), when $0 \leqslant \alpha \leqslant \pi$, while it is the diameter ($d=2r$), when $\pi < \alpha \leqslant 2\pi$. The ratio of the arc height, $h = r\left( 1 - \cos \frac{\alpha}{2}\right)$, to the maximal dimension of the arc is the reciprocal aspect ratio
\begin{equation}\label{eq:AR}
  \varepsilon^{-1} =
  \begin{cases}
    \frac{1}{2}\tan\frac{\alpha}{4}, & \text{if } 0 \leqslant \alpha \leqslant \pi ,\\
    \sin^2 \frac{\alpha}{4}, & \text{if } \pi < \alpha \leqslant 2\pi.
  \end{cases}
\end{equation}
The reciprocal aspect ratio varies from 0 (linear segment or stick) to 1 (circle).

\begin{figure}[!htb]
  \centering
  \includegraphics[width=0.8\columnwidth]{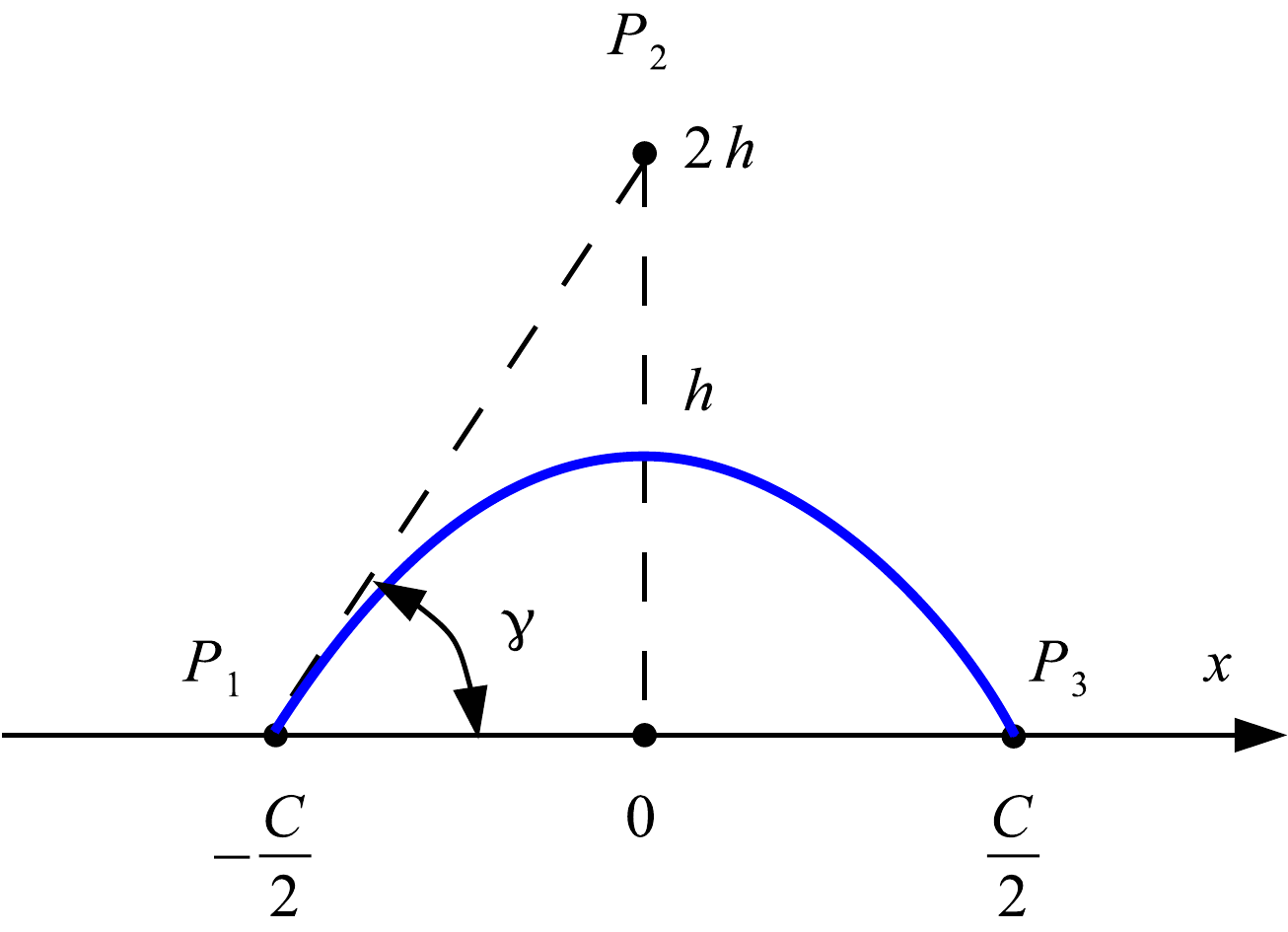}\\
  \caption{A symmetric quadratic B\'{e}zier curve.\label{fig:Beziercurve}}
\end{figure}

When the curved NW is mimicked using a symmetric quadratic B\'{e}zier curve of  unit length~\cite{Lee2021}, its shape is governed by a sole parameter, i.e., the angle, $\gamma$ (Fig.~\ref{fig:Beziercurve}). When $\gamma=0$, the  B\'{e}zier curve is a segment of unit length. When $\gamma=\pi/2$, the  B\'{e}zier curve is a unit length segment folded in half, i.e., a half-length segment (stick). Although one limiting case for the circular arcs and the symmetric quadratic B\'{e}zier curves coincides, the opposite limiting case differs for these two shapes. The reciprocal aspect ratio of the symmetric quadratic B\'{e}zier curve
\begin{equation}\label{eq:ARB}
\varepsilon^{-1} = \frac{1}{4} \tan \gamma
\end{equation}
varies from 0 (linear segment of unit length) to $\infty$ (half-length segment). The maximal dimension of the symmetric quadratic B\'{e}zier curve is the chord length, $C$, when $\tan\gamma <2$, while it is the chord height, $h$, otherwise.

\subsection{Simulation}\label{subsec:simul}
To reduce the effect of boundaries during deposition of the arcs, periodic boundary conditions (PBCs) were applied along both mutually perpendicular directions of the domain. Arcs were added one by one randomly, uniformly, and isotropically onto the  substrate, until a  cluster wrapping around the domain (torus) in two directions had arisen. Since our goal was to catch trends, rather than carry out a precise determination of the percolation threshold, we did not perform a finite-size analysis and  used only one, fixed, system size ($L=32$). To check for any occurrences of wrapping clusters, we used the version of the union--find algorithm~\cite{Newman2000PRL,Newman2001PRE} adapted for continuous percolation~\cite{Li2009PRE,Mertens2012}.

When calculating the electrical conductivity, the periodic boundary conditions were removed after reaching a pre-determinated value of the number density of conductive arcs, and then a potential difference was applied to the opposite boundaries of the system. We used the so-called multi-nodal representation~\cite{Rocha2015}, i.e., both the junction resistance and the resistance of the wires were taken into account. In such a way, a random resistor network (RRN) was generated. The structure of this RRN is irregular while its branches owing different resistances. In our study, the electrical resistance per unit length of the NWs was $\rho_\text{w}$, while the electrical resistance of a contact between any two nanowires (the junction resistance) was~$R_\text{j}$. Applying Ohm's law to each branch and Kirchhoff's point rule to each junction, we obtained a system of linear equations (SLEs). Although any such SLE can be huge, its numerical solution does not present significant difficulties since its matrix is sparse. We used the EIGEN library\cite{eigenweb} for this purpose.

\subsection{Mean-field approach}\label{subsec:analytics}
\begin{figure}[!htb]
  \centering
  \includegraphics[width=0.75\columnwidth]{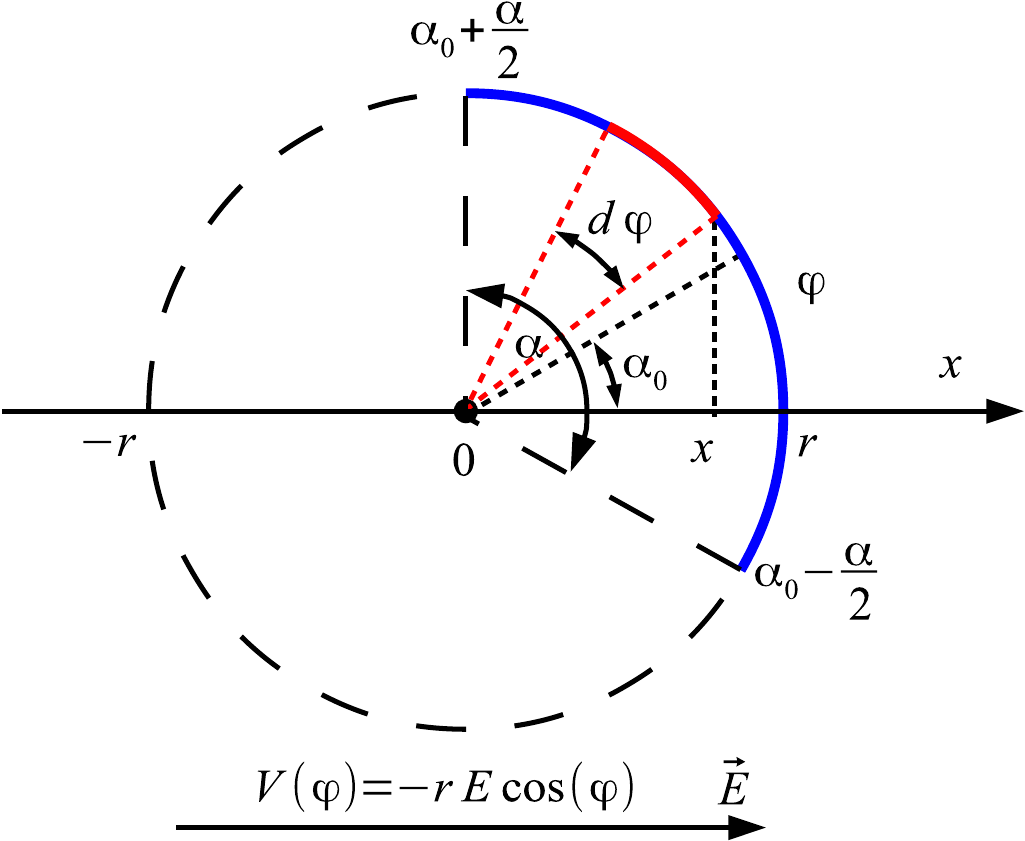}
  \caption{A conductive arc placed in an electrical field.\label{fig:arc}}
\end{figure}

The number density of the conductive NWs is assumed to exceed the percolation threshold $n_\text{c}$.
When a potential difference is applied to two opposite boundaries of the domain, the electrical potential varies almost linearly along the system~\cite{Sannicolo2018,Forro2018,Azani2019,Papanastasiou2021,Tarasevich2022}. The denser the system, the smaller is any deviation from the linearity. The main idea behind the mean-field approach is the consideration of a sole nanowire placed in the mean electrical field produced by all the rest of the nanowires, rather than a consideration of the complete ensemble of nanowires. To efficiently account for the contacts between the reference nanowire and the other nanowires, it is convenient to introduce the idea of a leakage conductance. Let a conductive arc be placed in a coordinate-dependent electrical field, $V(x)$ (Fig.~\ref{fig:arc}).

The orientation of the arc relative to the field is described by an angle $\alpha_0$. Now, let this arc be covered by an insulator that is characterized by a leakage conductance per unit length of the conductor. This leakage conductance is proportional to the average number of contacts, $\langle k \rangle$,
\begin{equation}\label{eq:G}
  G = \frac{\langle k \rangle}{\alpha r R_\text{j}}.
\end{equation}

Further consideration exactly corresponds to that presented in Ref.~\onlinecite{Tarasevich2021} for the case of conducting rings.
Let us consider the conductor between the points with angular coordinates $\varphi$ and $\varphi + d\varphi$.  According to Ohm's law,
\begin{equation}\label{eq:dudphi}
\frac{d  u(\varphi)}{d  \varphi} + i(\varphi)r \rho_\text{w}  = 0.
\end{equation}
Due to the conservation of electric charge,
\begin{equation}\label{eq:didphi}
\frac{d i (\varphi)}{d \varphi}  + [u(\varphi) - V(\varphi)] r G = 0.
\end{equation}
These two first-order ordinary differential equations (ODEs) can be combined in one second-order ODE
\begin{equation}\label{eq:d2udphi2}
\frac{d^2 u(\varphi)}{d \varphi^2} - \lambda^2 [u(\varphi) - V(\varphi)] = 0, \quad \lambda^2=r^2 \rho_\text{w} G.
\end{equation}
The solution of this equation is
\begin{equation}\label{eq:solu}
  u(\varphi) = A_1 \exp(\lambda \varphi) + A_2 \exp(-\lambda \varphi) - \frac{\lambda^2 r E\cos \varphi}{\lambda^2 +1}.
\end{equation}
Using~\eqref{eq:dudphi}, the electrical current can be written as follows
\begin{multline}\label{eq:soli}
i(\varphi) =  - \frac{\lambda }{r\rho_\text{w}} \left[ A_1 \exp(\lambda \varphi) - A_2 \exp(-\lambda \varphi)\right]\\ - \frac{\lambda^2 E}{\rho_\text{w}(\lambda^2 +1)} \sin \varphi.
\end{multline}
The coefficients $ A_1$ and $A_2$ can be found using the fact that, at the ends of the conductor, the electric current has to be zero
\begin{equation}\label{eq:BCi}
  i\left(\alpha_0 - \frac{\alpha}{2} \right) =i\left(\alpha_0 + \frac{\alpha}{2} \right) = 0.
\end{equation}
Thus,
\begin{multline}\label{eq:ivsalphaalpha0}
i(\varphi;\alpha_0,\alpha) =  \frac{\lambda^2 E}{\rho_\text{w}(\lambda^2 +1)}\\
\times\left\{
\frac{
\sin\left( \alpha_0 +\frac{\alpha}{2} \right) \sinh\left[\lambda \left(\varphi-\alpha_0+\frac{\alpha}{2}\right)\right]}
{\sinh(\lambda \alpha)}\right. \\ -
\left.\frac{\sin\left( \alpha_0 -\frac{\alpha}{2} \right) \sinh\left[\lambda \left(\varphi-\alpha_0-\frac{\alpha}{2}\right)\right]
}{\sinh(\lambda \alpha)} -  \sin \varphi
\right\}.
\end{multline}
When the arcs are oriented symmetrically with respect to the direction of the external electrostatic field ($\alpha_0=0$),
\begin{multline}\label{eq:ivsalpha}
i(\varphi;0,\alpha) =  \frac{\lambda^2 E}{\rho_\text{w}(\lambda^2 +1)}\\ \times
\left[
\frac{
\sin\left(\frac{\alpha}{2} \right) \sinh(\lambda\varphi)
}
{\sinh\frac{\lambda \alpha}{2}} -  \sin \varphi
\right].
\end{multline}
When additionally $\alpha = 2\pi$ (a complete circle), \eqref{eq:ivsalpha} reduces to the published formula~\cite{Tarasevich2021} for rings
\begin{equation}\label{eq:iring}
  i(\varphi;0,2\pi) =  -\frac{\lambda^2 E}{\rho_\text{w}(\lambda^2 +1)}\sin \varphi.
\end{equation}

For convenience, we will further consider rings consisting of an arc $\alpha$ with electrical conductivity per unit length $\rho_\text{w}$ and an arc $2\pi-\alpha$ with zero electrical conductivity. Obviously,  electric current can only flow in that part of the ring that has non-zero electrical conductivity, i.e., in the arc $\alpha$. The average current at some point of the ring can be obtained by integrating over all admissible orientations of the conducting arc of the ring.
The average current
\begin{equation}\label{eq:meani}
  \langle i(\varphi;\alpha)\rangle = \frac{1}{2\pi} \int \limits_{\varphi-\frac{\alpha}{2}}^{\varphi+\frac{\alpha}{2}} i(\varphi;\alpha_0,\alpha) \, \mathrm{d}\alpha_0
\end{equation}
is
\begin{multline}\label{eq:meaniphi}
\langle i(\varphi;\alpha)\rangle = \frac{\lambda^2 E}{\pi\rho_\text{w}(\lambda^2 +1)}\\ \times
\left[
\frac{\lambda}{ \left(\lambda^2 + 1\right)}
 \frac{\cosh(\lambda\alpha)- \cos\alpha}{\sinh(\lambda\alpha)} - \frac{\alpha}{2}
\right]\sin \varphi
\end{multline}
or in Cartesian coordinates
\begin{multline}\label{eq:meanix}
\langle i(x;\alpha)\rangle = \frac{\lambda^2 E}{\pi\rho_\text{w}(\lambda^2 +1)} \sqrt{1-\left(\frac{x}{r}\right)^2}\\ \times
\left[\frac{\alpha}{2} -
\frac{\lambda}{ \left(\lambda^2 + 1\right)}
 \frac{\cosh(\lambda\alpha)- \cos\alpha}{\sinh(\lambda\alpha)}
 \right].
\end{multline}
The total current that flows through all the arcs crossing any given equipotential is
\begin{equation}\label{eq:totalIint}
  I = 2 n L \int\limits_{-r}^{r} \langle i(x;\alpha)\rangle  \, \mathrm{d}x.
\end{equation}
Hence,
\begin{equation}\label{eq:totalcurrent}
  I = \left[
\frac{\lambda}{\lambda^2 + 1} \frac{\cosh(\lambda\alpha)- \cos\alpha}{\sinh(\lambda\alpha)} - \frac{\alpha}{2}
\right]
\frac{ n r \lambda^2 E L}{\rho_\text{w}(\lambda^2 +1)}.
\end{equation}
Thus, the electrical conductance is
\begin{equation}\label{eq:conductance}
  \sigma = \left[
  \frac{1}{2} -
\frac{\lambda}{\lambda^2 + 1} \frac{\cosh(\lambda\alpha)- \cos\alpha}{\alpha\sinh(\lambda\alpha)}
\right]\frac{ n l_\text{w} \lambda^2}{ \rho_\text{w}(\lambda^2 +1)}.
\end{equation}
The average number of contacts per arc
\begin{equation}\label{eq:kmean0}
  \langle k \rangle =  \frac{2 n l_\text{w}^2}{\pi}
\end{equation}
can either be written using the idea proposed in Ref.~\onlinecite{Yi2004} or derived explicitly. Since the derivation of the average number of contacts is very cumbersome, it is presented as Supplemental Material~\footnote{See Supplemental Material at [URL will be inserted by publisher] for the derivation of the average number of contacts between arcs.}. Therefore, the average leakage conductance per unit length is
\begin{equation}\label{eq:Gpul}
G = \frac{2 n  l_\text{w}}{\pi R_\text{j}}.
\end{equation}
Hence,
\begin{equation}\label{eq:lambda2subs}
\lambda^2 =   \frac{2 n l_\text{w}^3 \rho_\text{w}}{\alpha^2 \pi R_\text{j}}.
\end{equation}

\section{Results}\label{sec:results}
\subsection{Percolation threshold\label{subsec:percolation}}

\begin{figure}[!htb]
  \centering
  \includegraphics[width=\columnwidth]{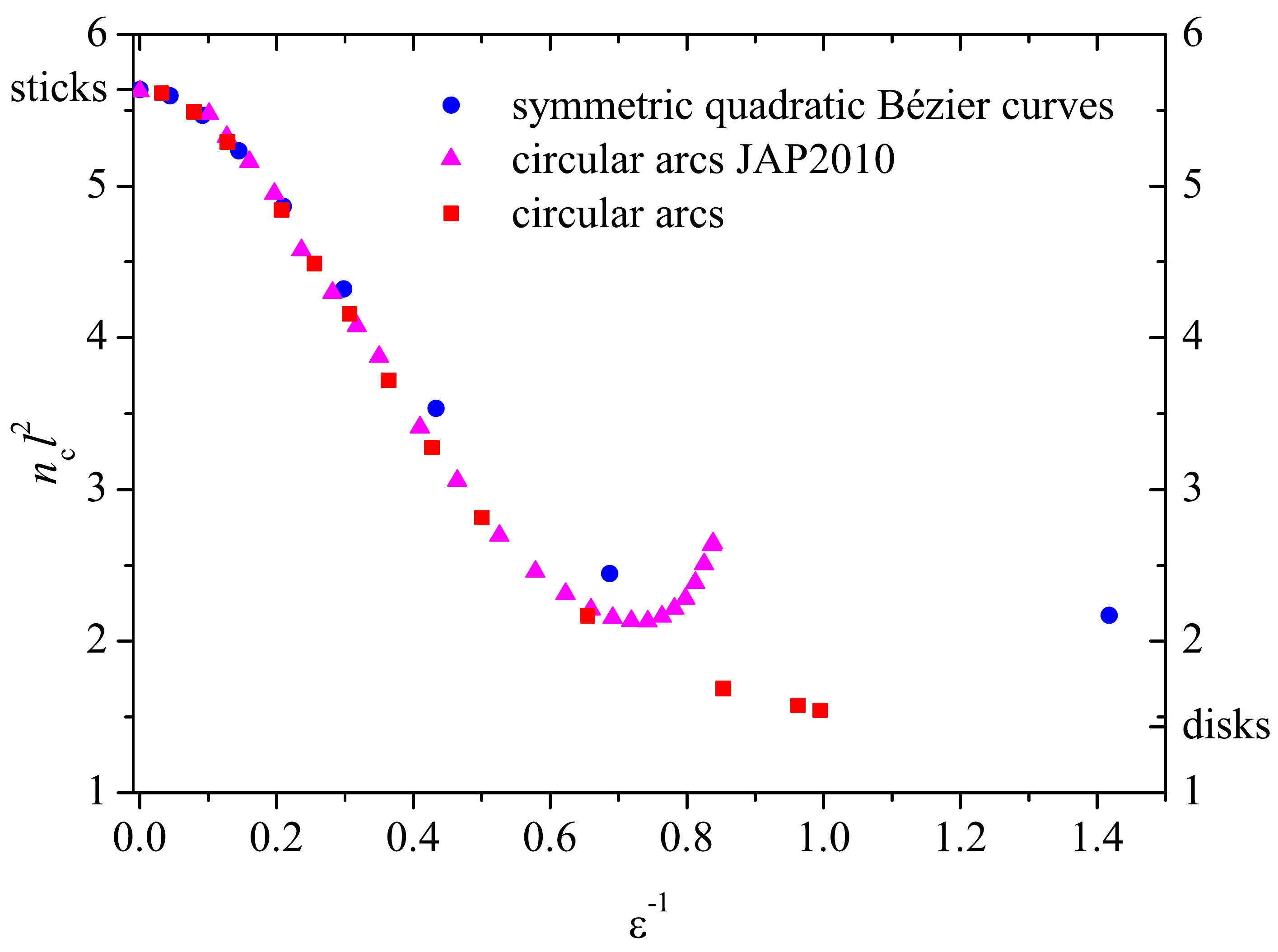}
  \caption{The percolation threshold, $n_\text{c} l^2$, is plotted against the reciprocal aspect ratio, $\varepsilon^{-1}$. The data for the symmetric quadratic B\'{e}zier curves (circles) are adapted from Ref.~\onlinecite{Lee2021} (see also Table~\ref{tab:Bezier}). Triangles correspond to the data  for circular arcs adapted from Ref.~\onlinecite{Lin2010} (see also Table~\ref{tab:arcs}).  Our simulation for circular arcs is shown as boxes. Here, $l$ is the maximal dimension of the particle.\label{fig:percolation}}
\end{figure}

Figure~\ref{fig:percolation} compares the dependencies of the percolation threshold, $n_\text{c} l^2$, on the value of the aspect ratio for curved wires of the particular shapes, viz., the circular arcs (our results along with published data~\cite{Lin2010}) and symmetric quadratic B\'{e}zier curves (adapted from Ref.~\onlinecite{Lee2021}). Here, $l$ means the maximal dimension of the particle.

\begin{table}[!htbp]
\caption{The percolation threshold for symmetric quadratic B\'{e}zier curves. The values of $\gamma$ and $n_\text{c}$ were taken from Ref.~\onlinecite{Lee2021}, while other values have been calculated by us. \label{tab:Bezier}}
\begin{ruledtabular}
\begin{tabular}{cccccc}
$\gamma$ & $\varepsilon^{-1}$ & $C$ & $h$ & $n_\text{c}$ & $n_\text{c}l^2$ \\
  \hline
$10^\circ$ & 0.04408 & 0.99487 & 0.04386 & 5.6542 & 5.59632 \\
$20^\circ$ & 0.09099 & 0.97880 & 0.08906 & 5.7065 & 5.46710 \\
$30^\circ$ & 0.14434 & 0.94961 & 0.13706 & 5.8042 & 5.23400 \\
$40^\circ$ & 0.20977 & 0.90309 & 0.18945 & 5.9694 & 4.86852 \\
$50^\circ$ & 0.29794 & 0.83202 & 0.24789 & 6.2457 & 4.32363 \\
$60^\circ$ & 0.43301 & 0.72455 & 0.31374 & 6.7358 & 3.53608 \\
$70^\circ$ & 0.68687 & 0.56252 & 0.38638 & 7.7344 & 2.44737 \\
$80^\circ$ & 1.41782 & 0.32319 & 0.45822 & 10.3331& 2.16962 \\
\end{tabular}
\end{ruledtabular}
\end{table}

The percolation threshold decreases as the aspect ratio increases. Up to $\varepsilon \approx 0.25$, the percolation threshold is insensitive to the particular shape of the curved wires. For larger values of the aspect ratio ($\varepsilon \gtrapprox 0.5$), the difference between the dependencies is more pronounced. This effect is not unexpected, since, as the aspect ratio increases, circular arcs tend to circles, while   symmetric quadratic B\'{e}zier curves first tend to U-shaped curves, and then to half-length line segments (see Section~\ref{subsec:sampling}). This behavior suggests that, when the percolation of slightly curved wires is simulated, the particular choice of a curve to mimic the shape of the real-world wires is hardly of the great importance. Circular arcs seem to be the simpler and more convenient choice as compared to   B\'{e}zier curves. However, comparison of our results with published data for arcs~\cite{Lin2010} (see also Table~\ref{tab:arcs}) suggests that those data~\cite{Lin2010} are hardly reliable for $\varepsilon^{-1} > 0.6$.
\begin{table}[!htbp]
\caption{The percolation threshold for arcs. The values in the first and second columns have been taken from Ref.~\onlinecite{Lin2010} [we digitalized the data presented in Fig.~8(b)~\cite{Lin2010}], while other values were calculated by us. \label{tab:arcs}}
\begin{ruledtabular}
\begin{tabular}{ccccc}
$\varkappa$ & $\phi_\text{c}$ & $\alpha$ & $\varepsilon^{-1}$  & $n_\text{c}l^2$ \\
  \hline
1.0000 & 0.562605 & 0.000 & 0 & 5.62605 \\
1.0275 & 0.570856 & 0.805 & 0.10132 & 5.47615 \\
1.0425 & 0.579107 & 0.995 & 0.12711 & 5.32341 \\
1.0668 & 0.587361 & 1.238 & 0.15988 & 5.15913 \\
1.1000 & 0.598915 & 1.498 & 0.19652 & 4.94971 \\
1.1400 & 0.605531 & 1.750 & 0.23592 & 4.57868 \\
1.2000 & 0.618742 & 2.053 & 0.2819 & 4.29682 \\
1.2500 & 0.636900 & 2.262 & 0.31735 & 4.07616 \\
1.3000 & 0.654848 & 2.443 & 0.35002 & 3.87484 \\
1.4000 & 0.668306 & 2.745 & 0.40956 & 3.40972 \\
1.5000 & 0.688141 & 2.992 & 0.46383 & 3.0584 \\
1.6250 & 0.709640 & 3.246 & 0.52599 & 2.69468 \\
1.7503 & 0.734434 & 3.457 & 0.57856 & 2.458 \\
1.8750 & 0.764170 & 3.636 & 0.62235 & 2.31208 \\
2.0000 & 0.793906 & 3.791 & 0.65951 & 2.20965 \\
2.1250 & 0.830232 & 3.927 & 0.69128 & 2.15378 \\
2.2500 & 0.873150 & 4.047 & 0.71867 & 2.13262 \\
2.3750 & 0.919360 & 4.154 & 0.74246 & 2.131 \\
2.5000 & 0.977103 & 4.251 & 0.76328 & 2.16312 \\
2.6252 & 1.04144 & 4.338 & 0.78163 & 2.21341 \\
2.7500 & 1.11236 & 4.418 & 0.79783 & 2.2798 \\
2.8773 & 1.20305 & 4.492 & 0.81252 & 2.3849 \\
3.0000 & 1.30198 & 4.558 & 0.82518 & 2.50708 \\
3.1264 & 1.40585 & 4.620 & 0.8369 & 2.63429 \\
3.1416 & 1.41575 & 4.627 & 0.83823 & 2.64459 \\
  \end{tabular}
  \end{ruledtabular}
  \end{table}

\subsection{Electrical conduction\label{subsec:conductivity}}

\begin{figure}[!htb]
  \centering
  \includegraphics[width=\columnwidth]{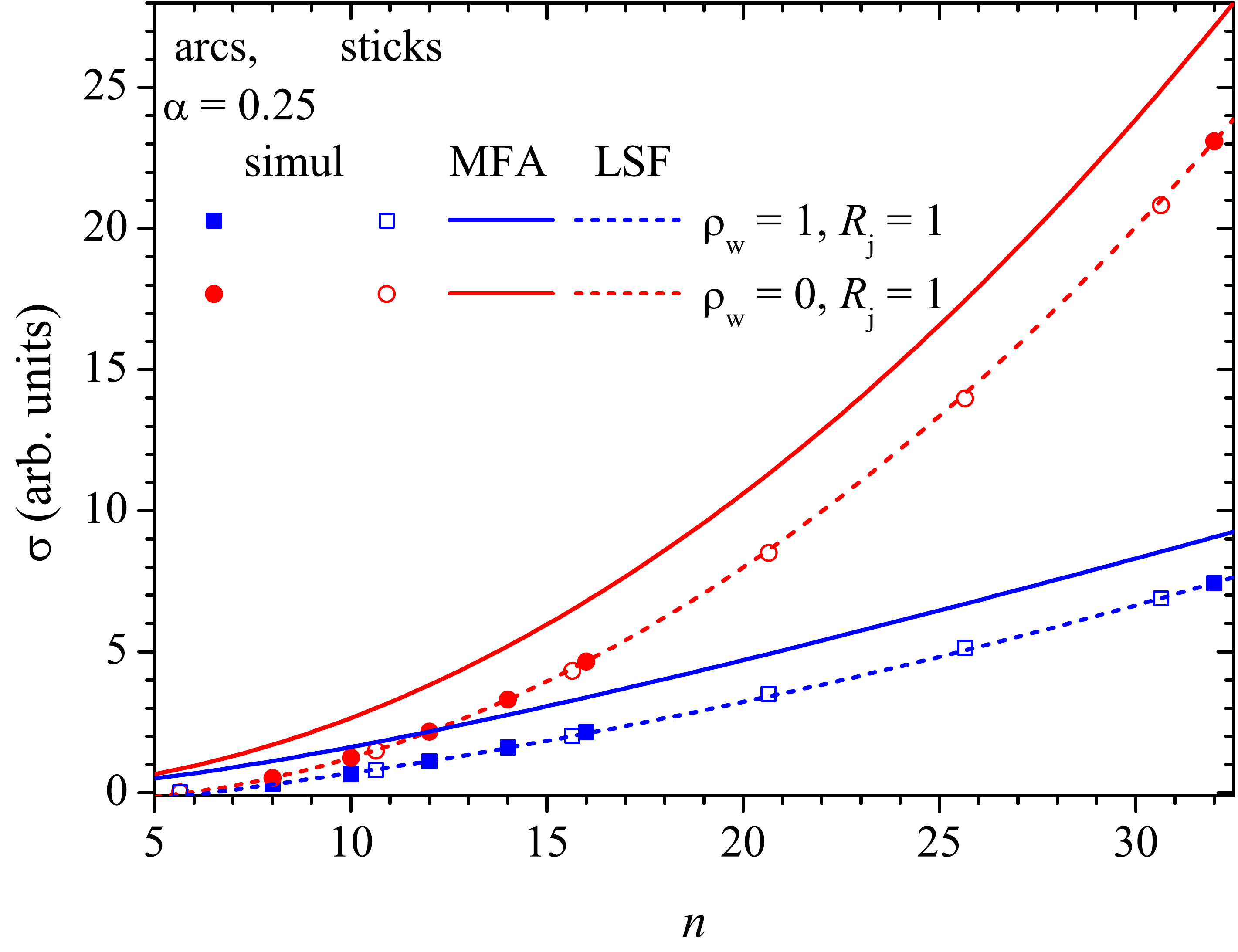}
  \caption{Electrical conductance plotted against the number density of conductive fillers. Open markers correspond to published data for random 2D systems of sticks~\cite{Tarasevich2022}, while filled markers correspond to random 2D systems of short arcs. Dashed lines represent least square fits using second-order polynomials, while solid lines demonstrate the MFA predictions for sticks.\label{fig:cond-vs-n}}
\end{figure}

To validate our computations of the electrical conductance, we compared the electrical conductance for systems of small arcs with the published results for sticks~\cite{Tarasevich2022}. When the curvature radius tends to infinity, i.e., the central angle tends to 0, the arc tends to a linear segment (stick). Figure~\ref{fig:cond-vs-n} demonstrates the electrical conductance plotted against the number density of conductive fillers.

Open markers correspond to published data for random 2D systems of sticks~\cite{Tarasevich2022}, while filled markers correspond to random 2D systems of short arcs ($r=4$, $l_\text{w}=1$, i.e., $\alpha = 0.25$). The dashed lines represent least square fits (LSFs) using second-order polynomials, while solid lines demonstrate the MFA predictions for sticks. When the junction resistance dominates over the wire resistance, the MFA predicts
\begin{equation}\label{eq:MFAJDA}
  \sigma = \frac{ l_\text{w}^4}{12 \pi R_\text{j}}n^2,
\end{equation}
that is $\sigma \approx 0.0265 n^2$ for the value of $l_\text{w}$ and $R_\text{j}$ used in our computations.
For comparison, the LSF gives $\sigma \propto 0.0263(2) n^2 $. When the junction resistance and the wire resistance are equal, the MFA predicts that for intermediate values of $n$~\cite{Tarasevich2022}
\begin{equation}\label{eq:MFAJWR}
  \sigma = \frac{ n l_\text{w}^2}{2 R_\text{j}}\left[ 1 - \frac{1}{l_\text{w}} \sqrt{\frac{2\pi}{n}} \tanh\left( l_\text{w} \sqrt{\frac{n}{2\pi}} \right) \right].
\end{equation}
In this situation, the LSF gives $\sigma \propto 0.0072(2) n^2 $. This test demonstrates that our data for arcs with small curvature are consistent with the published data for sticks (zero curvature).

\begin{figure}[!htb]
  \centering
  \includegraphics[width=\columnwidth]{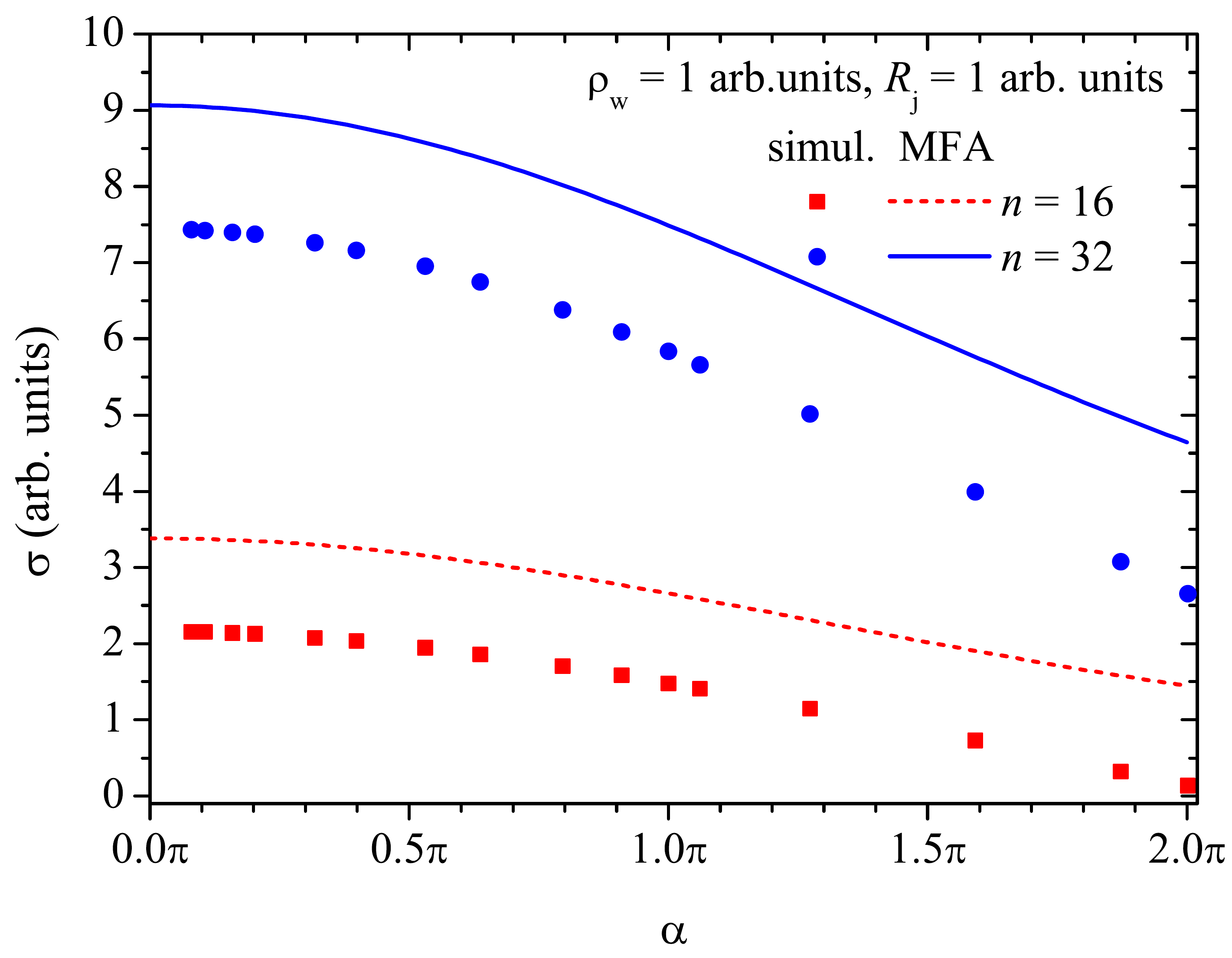}
  \caption{Dependence of the electrical conductance on the arc angle for two different values of the number density. The junction resistance and the wire resistance are equal. \label{fig:JWR}}
\end{figure}

Figure~\ref{fig:JWR} compares the prediction of the MFA and the results of computer simulation when the junction resistance, $R_\text{j}$, and the wire resistance, $R_\text{w} = l_\text{w} R_\text{w}$, are equal. Here, there is monotonically decreasing dependency of the electrical conductance on the arc size.

\begin{figure}[!htb]
  \centering
  \includegraphics[width=\columnwidth]{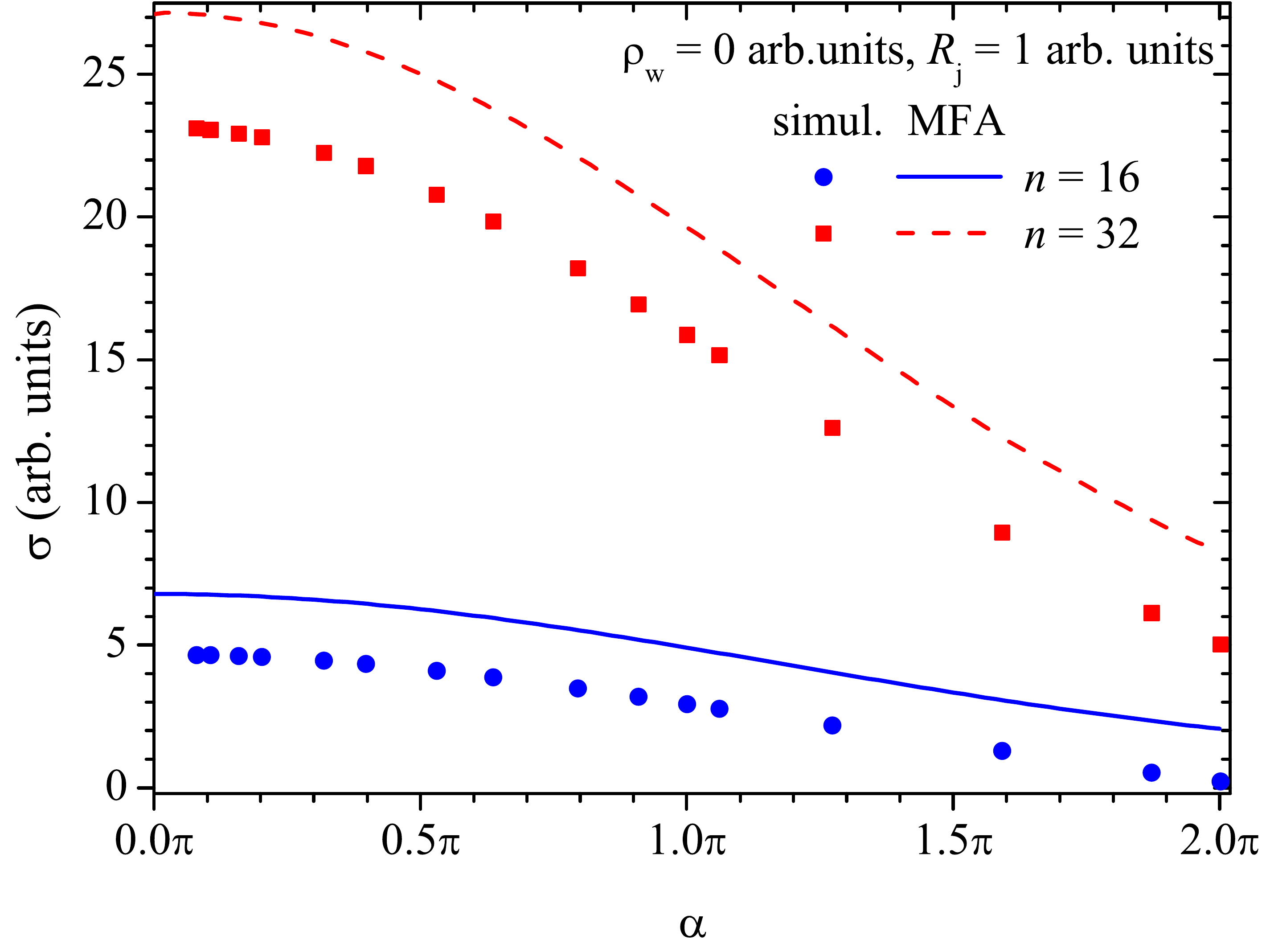}
  \caption{Dependence of the electrical conductance on the arc angle for two different values of the number density. The junction resistance dominates over the wire resistance. \label{fig:JDR}}
\end{figure}

Figure~\ref{fig:JDR} compares the prediction of the MFA and the results of computer simulations when the junction resistance dominates over the wire resistance. Again, there is monotonic dependency of the electrical conductance on the arc size. As the arc size increases, the electrical conductance  decreases.

\begin{figure*}
  \centering
  \includegraphics[width=\textwidth]{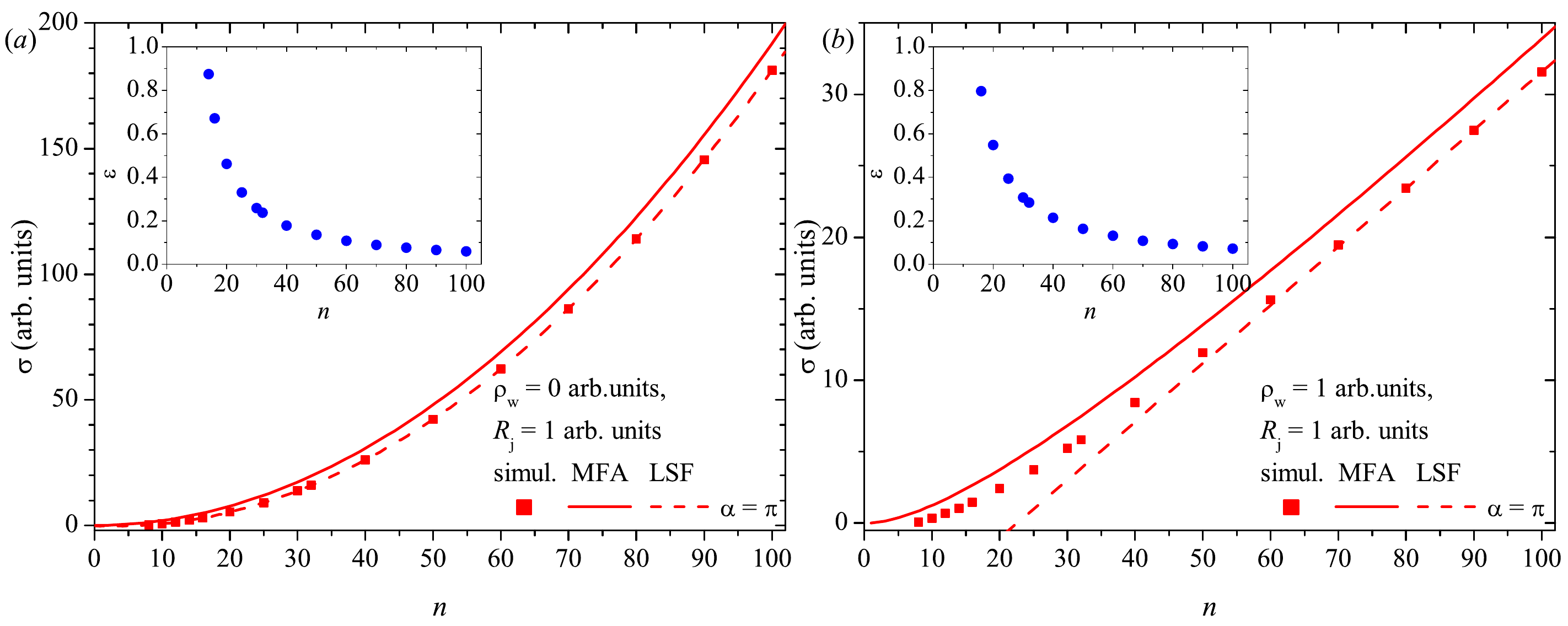}
  \caption{Examples of dependencies of the electrical conductance on the number density of conductive wires (semicircles, i.e., $\alpha = \pi$). (a) When the junction resistance dominates over the wire resistance. (b) When the junction resistance and the wire resistance are equal. Insets in both panels: the relative errors of the MFA predictions.}\label{fig:Semicircles}
\end{figure*}
As expected, the MFA overestimates the electrical conductance. The same behavior has been observed for random 2D systems of both conductive sticks~\cite{Tarasevich2022} and rings~\cite{Tarasevich2021}. However, as the number density of the curved conductive wires increases, the relative error of the MFA prediction decreases regardless of the ratios of the wire resistances to the junction resistances. Figure~\ref{fig:Semicircles} demonstrates the dependency of the electrical conductance on the number density of wires along with the relative errors of the MFA predictions for the case when the curved wires are semicircles, i.e., $\alpha = \pi$. When the junction resistance dominates over the wire resistance, the MFA predicts
\begin{equation}\label{eq:sigmaJDRsemicircles}
  \sigma =\frac{l_\text{w}^4 \left( \pi^2 - 4 \right) }{\pi^5 R_\text{j}} n^2.
\end{equation}
Since in our computations $l_\text{w} = 1$ and $R_\text{j}=1$, $\sigma \approx 0.01918 n^2$. The LSF  evidenced that $\sigma \propto 0.01928(6) n^2$.
When the junction resistance and the wire resistance are equal and  $n \gg 1$,
\begin{equation}\label{eq:sigmaJWRsemicircles}
  \sigma =\frac{l_\text{w} }{2 R_\text{j}} n.
\end{equation}
Since in our computations  $l_\text{w} =1$ and $R_\text{j} = 1$, $\sigma = 0.5 n$. The LSF evidenced that $\sigma \propto 0.408(2) n$.

The local (or apparent) conductivity exponent depends on the ratio $R_\text{j}/R_\text{w}$. Well above the percolation threshold ($n \gg n_\text{c}$), $t = 2$ when the junction resistance dominates over the wire resistance, while $t = 1 $ when the junction resistance and the wire resistance are equal. Similar results have been reported for random 2D systems of conductive sticks~\cite{Zezelj2012}.

\section{Conclusion}\label{sec:concl}
We have studied 2D random percolating networks of curved zero-width metallic nanowires. We mimicked the curved nanowires using circular arcs. Our computer simulations evidenced that the percolation threshold decreases as the curvature of the arcs increases. Comparison with published data on the percolation thresholds of  symmetric quadratic B\'{e}zier curves suggests that, when the percolation of slightly curved wires is simulated, the particular choice of curve to mimic the shape of the real-world wires is of little  importance.

Considering the electrical properties, we took into account both the nanowire resistance per unit length and the junction (nanowire/nanowire contact) resistance. Using a mean-field approximation, we derived the total electrical conductance  of the random nanowire-based networks as a function of their geometrical and physical parameters. The MFA predictions have been confirmed by our Monte Carlo numerical simulations. We found, that the electrical conductance monotonically decreases as the angular sizes of the arcs increases. Although the MFA overestimates the electrical conductance, the relative error of the MFA predictions decrease as the number density increases.

\begin{acknowledgments}
We would like to acknowledge funding from the Foundation for the Advancement of Theoretical Physics and Mathematics ``BASIS'', grant~20-1-1-8-1. The authors would also like to thank A.~Danilova for technical
assistance.
\end{acknowledgments}

\bibliography{arcs}

\end{document}